\documentstyle[12pt]{article}

\begin{document}
%\draft

\title{A Nonlinear Realized Approach of SU(2) Chiral\\ Symmetry
 Spontaneous Breaking and \\Properties of Nuclear Matter}

\author{{Xiao-fu L\"{u}$^{1,2}$, Bao-xi Sun$^{1,4}$, Yu-xin
Liu$^{3,1,5,}$\thanks{ Corresponding author} ,} \\
{Hua Guo$^{3}$, and En-guang Zhao$^{1,5}$ }  \\
\normalsize{$^{1}$Institute of Theoretical Physics, The Chinese
Academy of Sciences,}\\
\normalsize{P. O. Box 2735, Beijing 100080, China}\\
\normalsize{$^{2}$Department of Physics, Sichuan University, Chengdu 610064, China}\\
\normalsize{$^{3}$ School of Physics, Peking University, Beijing 100871, China}  \\
\normalsize{$^{4}$ Institute of High Energy Physics, The Chinese
Academy of Sciences,} \\
\normalsize{P.O. Box 918(4),Beijing 100039, China}  \\
\normalsize{$^{5}$Center of Theoretical Nuclear Physics,
National Laboratory of} \\
\normalsize{Heavy ion Accelerator,Lanzhou 730000, China} }
%\date{\today}

\maketitle

\begin{abstract}
A nonlinear realization of SU(2) chiral symmetry spontaneous
breaking approach is developed in the composite operator
formalism. A Lagrangian including quark, gluon and Goldstone boson
degrees of freedom of the chiral quark model is obtained from the
QCD Lagrangian. A way to link the chiral symmetry spontaneous
breaking formalism at hadron level and that at quark level is
predicted. too. The application to nuclear matter shows that the
approach is quite successful in describing the properties of
nuclear matter and the quark condensate in it.
\end{abstract}


\newpage

 It has been known that chiral symmetry and its spontaneous
breaking is the key points in understanding many features of the
nature, such as the generation of some particles in strong
interaction physics\cite{Wein96}, the chiral property of many
amino acids\cite{Salam91}, and so on. Moreover, it has been widely
used in guiding the design and synthesis of chemical compounds to
control and improve their functions\cite{NOSH01}. In low energy
strong interaction physics, Goldstone bosons appear as the chiral
symmetry is spontaneously broken. To carry out the constraint by
the appearance of Goldstone bosons, several realization formalisms
of the chiral symmetry spontaneous breaking ($\chi$SB) have been
developed \cite{Wein96,Furn935,Pap99}. Furthermore, Considering
the effective degrees of freedom to describing hadron structure,
we know that several formalisms have been developed. For example,
the one with constituent quarks and Goldstone bosons has been
shown to be successful in describing some properties of
hadrons\cite{CR85T97,LLZ97,Faes012} and hadron
spectroscopy\cite{Gloz96}, even nucleon-nucleon
interactions\cite{BS01}. The one including constituent quarks and
gluons are very powerful to describe hadron
properties\cite{GGI758,YF812,RS00}, too. Moreover, Manohar and
Georgi proposed that, besides the constituent quarks, both the
Goldstone bosons and gluons are necessary to describe hadron
structure\cite{MG84}. Such a scheme has also been widely used to
investigate hadron spectroscopy and hadron interactions(see, for
example, Refs.\cite{Faes94,ZY95} ). However, which effective
degrees of freedom have more sophisticated QCD foundation is still
a significant topic in hadron physics and low energy QCD. In the
sprit of $\chi$SB and the composite operator scheme\cite{SV92} of
the QCD, we propose, in this letter, a new approach to realize the
$\chi$SB nonlinearly not only in formalism but also in practical
calculation and obtain an effective Lagrangian including
constituent quarks, Goldstone bosons and gluons. Then some light
is shed on the effective degrees of freedom of hadron structure.
We propose also a link between the formalism of $\chi$SB at hadron
level and that at quark level. As an application, we take the
approach to describe the properties of nuclear matter. By
implementing the Hellmann-Feynman theorem\cite{CFG92}, we evaluate
the quark condensate in nuclear matter in the present formalism,
too.

At first, we discuss the linear infinitesimal transformation in
general
\begin{equation}
\phi'_{n}(x)~=~\phi_{n}+i\epsilon\sum_{m}t_{nm}^\alpha\phi_{m}(x)
\end{equation}
where $t^\alpha$ is a generator of the symmetry group of the
Lagrangian and $\phi_{n}$
 is a spin-zero boson field or spin-zero composite operators. The quantum
effective action is defined by
\begin{equation}
\Gamma [\phi] \equiv -\int d^{3}x\sum_{n}\phi^{n}J_{\phi_{n}}(x)+W[J_{\phi}]
\end{equation}
\par
This quantum effective action has the same symmetric property as
the Lagrangian. It gives that
\begin{equation}
\sum_{n,m}\int\frac{\delta\Gamma[\phi]}{\delta\phi_{n}} t_{nm}^{\alpha}\phi_{m}
(x)d^{4}x~=~0
\end{equation}
\par
Introducing an effective potential $V(\phi)$ as $ \Gamma [\phi] =
- v V[\phi]$, where $v$ is the  space-time volume, one can rewrite
Eq.(3) as
\begin{equation}
\sum_{n,m}\frac{\partial V(\phi)}{\partial\phi_{n}}
t_{nm}^{\alpha}\phi_{m} = 0,
\end{equation}
where $\phi$ is independent of $x$. Taking the differentiation
with respect to $\phi_{l}$, we obtain
\begin{equation}
\sum_{n}\frac{\partial V(\phi)}{\partial\phi_{n}} t_{nl}^{\alpha}+
\sum_{n,m}\frac{\partial^{2}V(\phi)}{\partial\phi_{n}\partial\phi_{l}}
t_{mn}^{\alpha}\phi_{m}~=~0
\end{equation}
The symmetry spontaneous breaking appears when $\frac{\partial
V(\bar\phi)}{\partial\bar\phi_{m}}=0$ and $\bar\phi_{m}\not=0$
where $\bar{\phi}_{m}=\langle 0 |\phi_{m}|0\rangle$. In accordance
with the condition of the symmetry spontaneous breaking, Eq.(5)
can be written as
\begin{equation}
\sum_{n,m}\left(\frac{\partial^{2}V(\phi)}{\partial\phi_{n}
\partial\phi_{l}} \right)_{\phi=\bar{\phi}} t_{nm}^{\alpha}
\bar{\phi}_{m} = 0\, .
\end{equation}
It is apparent that, if the symmetry is broken, $\sum_{m}
t_{nm}^{\alpha} \bar{\phi}_{m}$ should not identically equate to
zero. The massless eigenvectors of the mass matrix
$\frac{\partial^{2}V(\phi)} {\partial\phi_{n}
\partial\phi_{l}} $ span then a linear space in which the Goldstone
bosons lie and the remainder has to be perpendicular to it. This
constraint can lead various forms of the realization of the
$\chi$SB $SU_{L}(2)\times SU_{R}(2) \supset SU(2) $. In this paper
we shall discuss the nonlinear realization of $SU_{L}(2)\times
SU_{R}(2) \supset SU(2)$ on the quark level at first. Then we
propose a link between the $\chi$SB at quark level and that at
hadron level.

Constructing four composite operators in terms of the current
quark fields $u$ and $d$ as
\begin{equation}
\psi_{4} = \bar{q}q, ~~~~   \psi_{i}~=~i\bar{q}
\gamma_{5}\tau^{i}q
\end{equation}
where $\tau^{i}$ is the Pauli matrice, we can verify that the
transformation of $\psi_{n}$(n=1,2,3,4) under the chiral symmetry
transformation of the quarks $e^{i \gamma_{5} \epsilon^{i}
\tau^{i}}$ can be written as
\begin{equation}
\delta\psi_{4} = 2\epsilon^{i}\psi_{i}, ~~~~ \delta\psi_{i} =
-2\epsilon^{i} \psi_{4}
\end{equation}
where $\epsilon^{i}$ are the infinitesimal parameters.

Analyses in QCD sum rules\cite{RRY85} and calculations in lattice
QCD\cite{Neg99} have shown that the vacuum expectation value of
quarks (quark condensate) is not zero, i.e., $ \langle
0\left|\bar{q}(x)q(x)\right| 0\rangle\not= 0 $. It indicates that
the chiral symmetry $SU_{L}(2)\times SU_{R}(2)$ is spontaneously
broken and chiral condensate appears. As a consequence, the
$\psi_{n}$ has to be separated into two parts: one contains
Goldstone bosons and the other does not. This can be done by
writing
\begin{equation}
\psi_{n}~=~\sum_{m}(e^{2i\xi^{i}T^{i4}})_{nm}\tilde{\psi}_{m}
\end{equation}
where  $T^{\alpha\beta}$ are the generators of the four
dimensional rotation group. To make the $\tilde{\psi}_{n}$ do not
contain Goldstone bosons, the $\tilde{\psi}_{n}$ can be taken as
$\tilde{\psi}_{n}=(\tilde{\psi}_{4},0,0,0)$. In the infinitesimal
form, we have then
\begin{equation}
\delta\psi_{i}=2\xi^{i}\tilde{\psi}_{4}
\end{equation}
Comparing Eq.(10) with Eq.(8), we obtain the relation between the
constituent quarks and the current quarks as
\begin{equation}
\tilde{q} = e^{-i \gamma_{5} {\tau}^{i} \xi^{i} } q
\end{equation}

In order to simplify the factor $e^{-i \gamma_{5} \xi^{i}
\tau^{i}}$ we introduce the Glodstone bosons as
\begin{equation}
\xi^{i}=(\tan^{-1}{\left|\vec{\zeta}\right|})\frac{\zeta^{i}}
{\left|\vec{\zeta}\right|}
\end{equation}
Using Eq.(12), we can obtain the nonlinear realization of the
$\chi$SB as
\begin{equation}
e^{-i\gamma_{5} (\tan^{-1}{\left|\vec{\zeta}\right|} ) \vec{\tau}
\cdot \vec{\zeta}/\vert \vec{\zeta} \vert } = \frac{1-i\gamma_{5}
\vec{\tau}\cdot \vec{\zeta}} {\sqrt{1+\vec{\zeta}^{2}}} \, .
\end{equation}
%where $\vec{t} = \frac{1}{2}\vec{\tau}$.

Since the theory leaves the SU(2) invariant, the $\tilde{q}$
constructs a representation of SU(2).  Then it is necessary to
have
\begin{equation}
e^{-i \gamma_{5} \vec{\epsilon} \cdot \vec{\tau} } e^{-i
\gamma_{5} (\tan^{-1} \vert \vec{\zeta} \vert ) \vec{\tau} \cdot
\vec{\zeta}/ \vert {\vec{\zeta}} \vert } = e^{-i \gamma_{5}
(\tan^{-1}\vert \vec{\zeta}^\prime \vert) \vec{\tau} \cdot
\vec{\zeta}^\prime/\vert \vec{\zeta}^{\prime} \vert }
e^{-i\vec{\tau} \cdot \vec{\theta}}
\end{equation}
After some algebraic calculation we obtain the relation between
the Goldstone bosons and the constituent quarks as
$$\displaylines{\hspace*{15mm} \delta \tilde{q} =
-i\vec{\tau}\cdot(\vec{\epsilon} \times {\vec{\zeta}})\, ,
\hfill{(15a)} \cr \hspace*{15mm} \vec{\theta} = \vec{\epsilon}
\times \vec{\zeta} \, , \hfill{(15b)} \cr \hspace*{15mm}
\delta\vec{\zeta} = \vec{\epsilon}(1- \vec{\zeta}^{2}) + 2
\vec{\zeta} (\vec{\epsilon} \cdot \vec{\zeta}) \, , \hfill{(15c)}
\cr }$$

Under these transformations, the QCD Lagrangian
%$$ {\cal{L}}_{QCD} = i \bar{q} \gamma ^{\mu} D _{\mu} q -
%\frac{1}{4} G_{\mu \nu}^{a} G^{a \mu \nu} $$
 can be effectively expressed in terms of the constituent quark, gluon
 and Goldstone bosons as
$$\displaylines{\hspace*{2mm}
{\cal{L}}_{eff}= \sum_{\tilde{q}} i \bar{\tilde{q}} \gamma^{\mu}
\left( \partial_{\mu} + \frac{i(\vec{\zeta} \times
\partial_{\mu} \vec{\zeta}) \cdot \vec{\tau} }{1+ \vec{\zeta}^{2}}  -
{\frac{i \gamma_{5} \vec{\tau} \cdot
\partial_{\mu}\vec{\zeta}} {1+\vec{\zeta}^{2} } } \right)
\tilde{q} + i \bar{\tilde{q}} \gamma^{\mu} \left( - i g
\frac{\lambda^{a}}{2} A^{a,p}_{\mu} \right) \tilde{q} \hfill{} \cr
\hspace*{15mm} - m \bar{\tilde{q}}\tilde{q} - \frac{1}{4}G_{\mu
\nu}^{a\,p} G^{a \mu \nu p} + {\cal{L}}_{M} \,  . \hfill{(16)}
\cr}$$
  where $m$ is the constituent quark mass resulting from
quark condensates. ${\cal{L}}_{M}$ is the Lagrangian of the
Goldstone boson fields $\vec{\zeta}$ and the $\tilde{\psi}_4$. In
this sense, the effective degrees of freedom to describe hadron
structure should involve constituent quarks, gluons and Goldstone
bosons. One can then recognize that the constituent quark model of
hadron structure involving constituent quarks, Goldstone bosons
and gluons\cite{MG84} has a quite solid QCD foundation.
Furthermore, the double counting and the spurious state involved
in Manohar-Georgi's formalism\cite{MG84} disappears in the present
approach.

>From the effective action $\Gamma$ in Eq.(2) we have
\setcounter{equation}{16}
 \begin{equation}
 {\cal{L}}_{M} = -\frac{1}{2A}\partial_{\mu}   \tilde{\psi}_{4}
  \partial^{\mu}\tilde{\psi}_{4}-\frac{2}{A}
 \tilde{\psi}_{4}\tilde{\psi}_{4}
 \vec{D}_{\mu}
 \vec{D}^{\mu} + {\cal{L}}(\tilde{\psi}_4  )
 \end{equation}
 where $D_{\mu}^{i} =\frac{\partial_{\mu}\zeta^{i}}{1+\vec{\zeta}^{2}}$.
$A$ is a dimensional constant. Considering the pseudoscalar field
property of $\zeta_{i}$ and the assumption of the saturation of
vacuum, we obtain that the ${\cal{L}}_{M}$ can be written as the
superposition of the following two parts
\begin{equation}
{\cal{L}}_{\sigma} =
-\frac{1}{2}\partial_\mu\sigma\partial^\mu\sigma - \frac{1}{2}
m^2_\sigma \sigma^2 - \frac{g_2}{3} \sigma^3 -
\frac{g_3}{4}\sigma^4_{} \, ,
\end{equation}
% Those expressions are just the same as those commonly used
%in chiral models. Meanwhile, we have the Lagrangian of pions (the
%Goldstone bosons) as
\begin{equation}
{\cal{L}}_{\pi} = - \frac{1}{2} \frac{\partial ^{\mu}
\vec{\pi}}{1+ \frac{\vec{\pi}^2}{f_{\pi}^2}} \cdot \frac{\partial
^{\mu} \vec{\pi}}{1+ \frac{\vec{\pi}^2}{f_{\pi}^2} } - 2 f_{\pi}
\sigma  \frac{\partial ^{\mu} \vec{\pi} \cdot \partial ^{\mu}
\vec{\pi} }{\left(1+ \frac{\vec{\pi}^2}{f_{\pi}^2} \right)^2 } - 2
\sigma ^2 \frac{\partial ^{\mu} \vec{\pi} \cdot \partial ^{\mu}
\vec{\pi}}{\left(1+ \frac{\vec{\pi}^2}{f_{\pi}^2} \right)^2 } \, .
\end{equation}
And the parameters $g_2$ and $g_3$ hold relation $g_2=-\frac{3 m_q
\sqrt{A}}{m_\sigma^2}g_3 $ with $m_q$ being the current quark
mass.

It has been well known that there exists $\chi$SB formalism at
hadron level. However the relation between the formalism and that
at quark level is not clear enough.  To link the $\chi$SB
formalism at hadron level with the $\chi$SB formalism at quark
level described above, we construct a set of composite operators
which have the quantum numbers of proton and neutron as
$$\displaylines{ \hspace*{10mm}
p(x) = \varepsilon_{abc}\left(u^{aT}(x)~,~d^{aT}(x) \right)
c\gamma^{\mu}u^{b}(x) \gamma_{5}\gamma_{\mu}i\tau_{2} \left(
\begin{array}{l}
 u^{c}(x)\\
 d^{c}(x)\\
\end{array}
\right),   \hfill{(20a)} \cr \hspace*{10mm}
 n(x) = \varepsilon_{abc}\left(u^{aT}(x)~,~d^{aT}(x) \right)
c\gamma^{\mu}d^{b}(x) \gamma_{5}\gamma_{\mu}i\tau_{2} \left(
\begin{array}{l}
 u^{c}(x)\\
 d^{c}(x)\\
\end{array}
\right),  \hfill{(20b)} \cr } $$
 where $a$, $b$, $c$ are the
color indices,  $\tau_2$ acts only on the isospin indices and the
row and column matrix express the doublet of the isospin. Since
the nucleon in this doublet are composed of current quarks, we can
refer them as current nucleon. Implementing the chiral
transformation of quarks $\tilde{q} =
\exp(-i\gamma_{5}\xi^{i}\tau^{i}) q$, we have the transition among
nucleons as   \setcounter{equation}{20}
\begin{equation}
\tilde{N} = e^{-i\gamma_{5}\xi^{i}\tau^{i}} N,
\end{equation}
where $  N  =  \left(
\begin{array}{l}
p(x)\\
n(x)\\
\end{array}
\right)  $ , $  \tilde{N}= \left(
\begin{array}{l}
\tilde{p}(x)\\
\tilde{n}(x)\\
\end{array}
\right) $ in which
$$\displaylines{ \hspace*{10mm}
\tilde{p}(x) = \varepsilon_{abc} \left(\tilde{u}^{aT}(x),
\tilde{d}^{aT}(x) \right) c\gamma^{\mu}\tilde{u}^{b}(x)
\gamma_{5}\gamma_{\mu}i\tau_{2} \left(
\begin{array}{l}
\tilde {u}^{c}(x)\\
\tilde {d}^{c}(x)\\
\end{array}
\right),  \hfill{(22a)}  \cr \hspace*{10mm}
\tilde{n}(x)=\varepsilon_{abc}\left(\tilde{u}^{aT}(x),
\tilde{d}^{aT}(x) \right) c\gamma^{\mu}\tilde{d}^{b}(x)
\gamma_{\mu}\gamma_{5}i\tau_{2} \left(
\begin{array}{l}
\tilde{u}^{c}(x)\\
\tilde{d}^{c}(x)\\
\end{array}
\right),  \hfill{(22b)} \cr }$$
 The Eq.~(21) shows that we can obtain the nucleon from the current
nucleon. The nucleon observed in experiments in the low energy
region can be formed only through Eq.~(22).

Taking the same measure as described for the quarks and
considering the Goldberg-Treiman relation and the partial
conservation of the axial current(PCAC), we obtain the Lagrangian
density with the nonlinear realization of the SU(2) $\chi$SB at
hadron level as\setcounter{equation}{22}
\begin{eqnarray}
{\cal L}_{ch}& = & \bar\psi \left(i\gamma_{\mu}\partial^{\mu}  -
M_N -\frac{\vec{\tau}\cdot\left(\vec{\pi}\times \rlap/{\partial}
\vec{\pi}\right)} {f_{\pi}^2 \left(1+\frac{\vec{\pi}^2}{
f_{\pi}^2}\right)} +\frac{2i M_N g_A \gamma_5 \vec{\tau}\cdot
\vec{\pi}} {f_{\pi}\left(1+\frac{\vec{\pi}^2}{f_{\pi}^2}\right)}
-\frac{2 g_A \gamma_5 \vec{\tau} \cdot \vec{\pi} \vec{\pi} \cdot
\rlap/{\partial} \vec{\pi}} {f_{\pi}^3 \left(1+\frac{\vec{\pi}^2}{
f_{\pi}^2}\right)^2}\right)\psi \nonumber \\
& & - \sigma^2 \bar\psi \left( \frac{8i g^{\prime} M_N \gamma_5
\vec{\tau}\cdot \vec{\pi}} {f_{\pi}
\left(1+\frac{\vec{\pi}^2}{f_{\pi}^2} \right)} +\frac{8 g^{\prime}
\gamma_5 \vec{\tau} \cdot \vec{\pi} \vec{\pi} \cdot
\rlap/{\partial} \vec{\pi}} {f_{\pi}^3 \left(1+\frac{\vec{\pi}^2}
{f_{\pi}^2} \right)^2}\right)\psi -\frac{\delta \! M_p + \delta \!
M_n}{2} \left(\frac{1-\frac{\vec{\pi}^2}{f_{\pi}^2}}
{1+\frac{\vec{\pi}^2}{f_{\pi}^2}} \right)\bar\psi\psi \nonumber\\
& & -\left(\delta \! M_p - \delta \! M_n \right) \left(
\frac{\tau_3}{2} - \frac{1}{f_{\pi}^2} \left( \frac{\pi_3}
{1+\frac{\vec{\pi}^2}{f_{\pi}^2}}\right) \vec{\tau} \cdot
\vec{\pi}\right)\bar\psi \psi
-g_\sigma\bar\psi\psi\sigma  \nonumber\\
& & - \frac{1}{2}\partial_\mu\sigma\partial^\mu\sigma - U(\sigma)
-\left( \frac{1}{2} + 2 f_{\pi} \sigma +2\sigma^2 \right)
\frac{\partial_\mu\vec{\pi} \cdot
\partial^\mu \vec{\pi}}{\left(1+\frac{\vec{\pi}^2}{f_{\pi}^2}
\right)^2 } - \frac{1}{2} m_{\pi}^2 \vec{\pi}^2 \, ,
\end{eqnarray}
where $M_N$ is the mass of the nucleon caused by the $\chi$SB,
$\delta \! M_p$ and $\delta \! M_n$ are the masses of the current
proton and neutron, respectively. The $U(\sigma )$ stands for
\begin{equation}
U\left(\sigma\right) = \frac{1}{2} m^2_\sigma \sigma^2_{} +
\frac{1}{3}g_2\sigma^3_{}
 + \frac{1}{4}g_3\sigma^4_{},
\end{equation}
where the $g_2$ and $g_3$ have relation $g_2=-\frac{3\delta
M_N\sqrt{A}}{m_\sigma^2}g_3 $ with $\delta M_N = \delta M_n =
\delta M_p$ .

To represent the repulsive interaction among nucleons and the
isospin symmetry breaking in nuclear matter, we can include
$\omega$ and $\rho$ mesons in the way having been widely
used\cite{SW97}. Then in the mean field approximation, we have the
Lagrangian
\begin{eqnarray}
{\cal L}_{RMF} & = & \bar\psi \left(i\gamma_{\mu}\partial^{\mu}  -
M_N \right)\psi
 - \delta \! M_N \left( \frac{1-\frac{\pi_{0}^2}{f_{\pi}^2}}
{1+\frac{\pi_{0}^2}{f_{\pi}^2}} \right) \bar\psi\psi \nonumber\\
& & - g_\sigma\bar\psi\psi\sigma_{0} -
g_\omega\bar\psi\gamma^{0}\psi\omega_{0}
- g_{\rho} \bar\psi \gamma^0 t _3 \rho_{03} \psi  \nonumber\\
& & -\frac{1}{2} m^2_\sigma \sigma^2_{0}-\frac{g_2}{3}\sigma^3_{0}
-\frac{g_3}{4}\sigma^4_{0} - \frac{1}{2} m_{\pi}^2 \pi_{0}^2
+\frac{1}{2} m^2_\omega\omega_{0}^2 + {1\over2}m_\rho^{2}
\rho_{03}^2 ,
\end{eqnarray}
where the $\sigma_0$ is the expectation value of the  isoscalar
scalar field, $\pi_{0}^{2}$ is that of $\vec{\pi}^{2}$, $\omega_0$
is that of the temporal component for the isoscalar vector field
since there is no spatial direction for a uniform nuclear matter
at rest, $\rho_{03}$ is that of $\rho$ meson in the nuclear
matter.

 By fitting the saturation properties of nuclear
matter, we obtain the parameters with the restriction $M_{N} +
\delta \! M_N = 938$~MeV. Two sets of the parameters are listed in
Table 1. For the parameter set $C_1$, we get the saturation
density of $0.152\mbox{fm}^{-3}$, binding energy of $15.297$~MeV,
a compression modulus of $349.10$~MeV, symmetry energy coefficient
$33.645$~MeV and the effective mass of a nucleon of $0.687M_N$ for
symmetric nuclear matter. The parameter set $C_2$ gives a
saturation density $0.151\mbox{fm}^{-3}$, a binding energy of
nucleon $15.040$MeV, a compression modulus $K=298.88$~MeV, a
symmetry energy coefficient $32.593$~MeV and an effective mass of
nucleon $0.736M_N$ for symmetric nuclear matter. Meanwhile the
curves for the equation of states (EOS) are obtained too. The
numerical results show that the EOSs for the two sets of
parameters are quite close to each other. we display then only the
equation of states for the parameter set $C_2$ in Fig. 1. The
figure shows evidently that the symmetric nuclear matter can exist
stably. However stable pure neutron matter does not exist.

Making use of the Hellmann-Feynman theorem\cite{CFG92}, one can
obtain the relation between the quark condensate in nuclear matter
$Q(\rho) = \langle 0\left|\bar{q}(x)q(x)\right| 0\rangle_{\rho \ne
0 } $ and that in vacuum  $ Q(0) = \langle
0\left|\bar{q}(x)q(x)\right| 0\rangle_{\rho=0}$ as
\begin{eqnarray}
\frac{Q(\rho)} {Q(0)} &=& 1 + \frac{1}{2Q(0)} \left(
\sum_{N=n,p}\frac{\partial
 \varepsilon}{\partial M_N} \frac{d M_N}{d m_q} +
\frac{\partial
 \varepsilon}{\partial M_{\sigma}} \frac{d M_{\sigma}}{d m_q} +
\frac{\partial \varepsilon}{\partial M_{\pi}} \frac{d M_{\pi}}
{dm_q} \right) \nonumber\\
 &=&  1 - \frac{2\sigma_N}{(m_{\pi} f_{\pi})^2}
\left(\rho_{s}(p)+\rho_{s}(n)\right)- \frac{\pi_0^2}{f_\pi^2}
-\frac{3g_3m_\pi^2\sigma_0^2}{2m_\sigma^4}
\end{eqnarray}
where $\sigma_N$ is the nucleon $\sigma$ term, $\rho_s = \langle
\bar{\psi} \psi \rangle $ is the scalar density of the nucleons.

With the parameter sets $C_2$ determined above and $\sigma_{N} =
45$~MeV, we evaluate the ratio of the in-medium quark condensate
to that in vacuum. The obtained results are represented in Fig.~2.
 From the figure one can easily know that the quark condensate in
nuclear matter decreases, in general, as the nuclear matter
density increases. It manifests that the ``upturn" problem in the
other approaches at hadron level\cite{LK94,MD97,Mit97} is removed
in the present $\chi$SB approach. Investigating the figure more
carefully, one may know that there exists a density $\rho= 0.235
\, \mbox{fm}^{-3}$ at which the decreasing rate is enhanced. Such
a behavior is consistent with the result given in
Ref.\cite{Bard90}. Relevant investigation shows that pion
condensate in strong interaction appears as the nuclear matter
density reaches $0.235\, \mbox{fm}^{-3}$ and beyond. It indicates
that the sudden increase of the decreasing rate or the
discontinuous decrease of the quark condensate in nuclear matter
may come from the appearance of pion condensate.

In summary, a nonlinear realization of SU(2) $\chi$SB approach is
developed in the composite operator formalism of QCD. A Lagrangian
including quark, gluon and Goldstone bosons of chiral quark model
is obtained from the QCD Lagrangian. It indicates that, besides
the constituent quarks, both the gluons and Goldstone bosons are
the effective degrees of freedom to describe hadrons. We
constructed also a way link the $\chi$SB at hadron level and that
at quark level. As an application, the properties of symmetric
nuclear matter and pure neutron matter are investigated in the
mean-field approximation. The calculation shows that the present
approach is quite powerful in describing the properties of nuclear
matter. Meanwhile the quark condensate is evaluated. It shows that
the quark condensate decreases with the increasing of nuclear
matter density. The chiral symmetry is then restored gradually in
nuclear matter as the density increases.

\bigskip

\bigskip

This work is supported by the National Natural Science Foundation
of China under contact No. 10047001, 10075002, 10135030, and
19975062, the Major State Basic Research Development Program under
contract No.G2000077400 and CAS Knowledge Innovation Project No.
KJCX2-N11. One of the author(Liu) thanks also supports by
Foundation for University Key Teacher by the Ministry of
Education, China. The authors are indebted to Professors F. Wang,
C. S. Chang, W. Q. Chao, P. F. Zhuang and H. S. Zong for their
helpful discussions.

\newpage
Table  1.  The parameters in the calculation of the SU(2) chiral
symmetry spontaneous breaking model with $m_\pi=139.57MeV$ and
$f_\pi=130MeV$ ($m_{i} (i=\sigma, \omega, \rho)$, $M_N$ and
$\delta \! M_N$ in Mev, $g_2$ in fm$^{-1}$).
\begin{center}
\begin{tabular}{|c|c|c|c|c|c|c|c|c|c|c|}\hline
 {}  & $g_{\sigma}(N)$ & $m_{\sigma}$ & $g_{\omega}(N)$ & $m_{\omega}$
    & $g_{\rho}(N)$ & $m_{\rho}$ & $g_2$
    & $g_3$ & $M_N$ & $\delta \! M_N$  \\ \hline
$C_1$ & 9.111 & 540 & 10.587 & 783 & 8.480 & 770  & $-4.0$
      & 20.0 & 888 & 50 \\  \hline
$C_2$ & 9.111 & 570 & 9.573 & 783 & 8.480 & 770 & $-9.0$
      & 37.5 & 888 & 50  \\ \hline
\end{tabular}
\end{center}

\newpage

\leftline{\Large {\bf Figure Captions}}
\parindent = 2 true cm
\parskip 1 cm
\begin{description}
\item[Fig.~1.]  Obtained average energy per nucleon $\varepsilon/\rho_N - M_N$
as a function of nucleon density $\rho_N$ for the parameter sets
$C_2$. The solid line denotes symmetric nuclear matter , and the
dashed line is for pure neutron matter.

\par

\item[Fig.~2.] Obtained ratio of the quark condensate in nuclear matter to
that in vacuum as a function of nucleon density for the parameter
sets $C_2$. The solid line denotes that in symmetric nuclear
matter, and the dashed line is that for pure neutron matter.

\end{description}

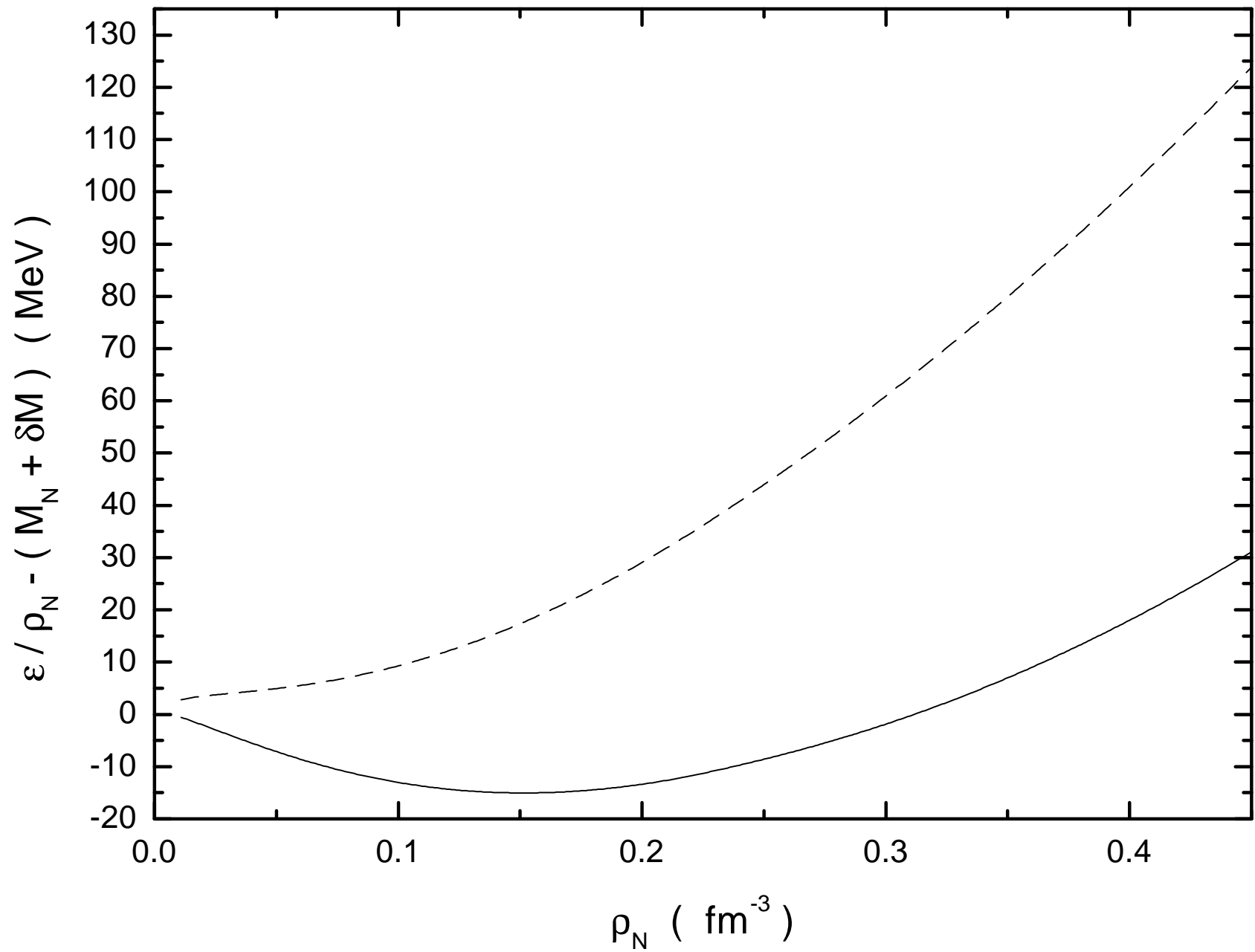
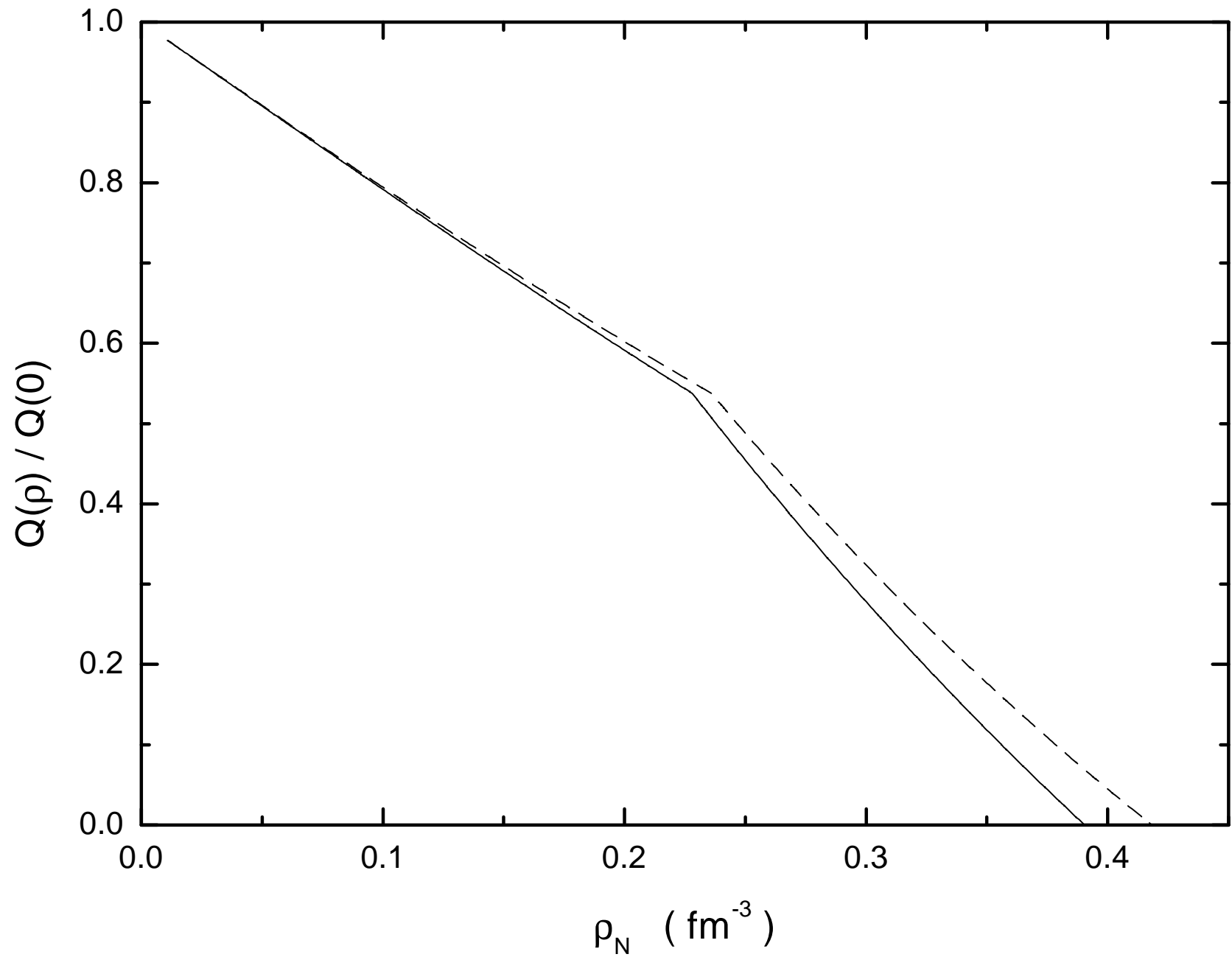
\end{document}